\documentclass[12pt]{article}

\newenvironment{numberedlist}
{\begin{list}{\makebox[20pt]{\hss(\arabic{itemno})\enspace}}
             {\usecounter{itemno}\labelwidth 20pt}}{\end{list}}

\newcounter{itemno}

\newcounter{itemno1}

\newcounter{itemno2}

\newcounter{exno}

\newcounter{defno}

\newenvironment{defn}{\refstepcounter{defno}\medskip \noindent {\bf
Definition \thedefno.\ }}{\medskip}

\newcommand{\sep}{\;\vert\;}

\newcommand{\oprove}{\vdash\kern-.6em\lower.7ex\hbox{$\scriptstyle O$}\,}

\newcommand{\Pscr}{{\cal P}}

\newcommand{\pderivation}{{\cal P}\kern -.1em\hbox{\rm -derivation}}
\newcommand{\pderivationl}{{\cal P}\kern -.1em\hbox{\em -derivation}}
\newcommand{\pderivable}{{\cal P}\kern -.1em\hbox{\rm -derivable}}
\newcommand{\pderivablel}{{\cal P}\kern -.1em\hbox{\em -derivable}}
\newcommand{\pderivations}{{\cal P}\kern -.1em\hbox{\rm -derivations}}
\newcommand{\pderivability}{{\cal P}\kern -.1em\hbox{\rm -derivability}}

\newcommand{\all}{\forall}
\newcommand{\some}{\exists}

\newsavebox{\lpartfig}
\newsavebox{\rpartfig}

\newenvironment{exmple}{
 \begingroup \begin{tabbing} \hspace{2em}\= \hspace{3em}\= \hspace{3em}\=
\hspace{3em}\= \hspace{3em}\= \hspace{3em}\= \kill}{
 \end{tabbing}\endgroup}

\newcommand{\lb}{\langle}
\newcommand{\rb}{\rangle}
\newcommand{\pr}{prov}

\newcommand{\prove}{exec} 
  
     {\\* \hspace*{\fill} \end{trivlist}}

\renewcommand{\pr}{pv}
\renewcommand{\prove}{ex} 

\begin{document}

\begin{center}
{\Large {\bf A New Execution Model for the logic of hereditary Harrop formulas}}
\\[20pt] 
{\bf Keehang Kwon}\\
Dept. of Computer  Engineering, DongA University \\
Busan 604-714, Korea\\
  khkwon@dau.ac.kr\\
\end{center}

\noindent {\bf Abstract}: 
The class of first-order Hereditary Harrop formulas ($fohh$) is a well-established extension of
first-order Horn clauses.
Its operational semantics is based on intuitionistic provability. 

   We propose another  operational
semantics for $fohh$ which is based on game semantics. This new semantics has several interesting
aspects: in particular, it gives a logical status to
the $read$ predicate.

{\bf keywords:} interaction, game semantics, read, computability logic.



\section{Introduction}\label{sec:intro}

The logic of first-order hereditary Harrop formulas is a well-established extension to the logic
of Horn clauses. Its operational semantics is based on intuitionistic provability.
In the operational semantics based on  provability such as  uniform provability 
\cite{Mil89jlp,MNPS91}, 
solving the universally quantified goal $\all x G$ from a program $\Pscr$
 simply {\it terminates} with a success if $[c/x] G$ is solvable from $\Pscr$ where $c$ is a new
constant.

 Our approach in this paper involves a modification of the  operational
semantics to allow for more active participation from the user.
Executing  $\all x G$ from a program $\Pscr$ now  has the
following two-step operational semantics:

\begin{itemize}

\item Step (1): the machine tries to prove $\all x G$ from a program $\Pscr$. If it fails, the
machine returns the failure. If it succeeds, goto Step (2).

\item Sstep (2): the machine requests the user to choose a constant $c$ for $x$ and then proceeds
with solving the  goal, $[c/x] G$.

It can be easily seen that our new semantics is more ``constructive'' than  the old semantics.
In particular, it gives a logical status to the $read$ predicate in Prolog.

\end{itemize}

As an   
illustration of this approach, let us consider the following program which computes the cube of a natural
number.
\begin{exmple}
$\all x \all y (cube(x,y) $ ${\rm :-}$ \> \hspace{7em}      $nat(x) \land y$ is $x*x*x)$\\ 
\end{exmple}
\noindent Here, ${\rm :-}$ represents reverse implication.
 As a particular example, consider a goal task $\all x \exists y (nat(x) \supset cube(x,y))$.
 This goal simply terminates
with a success in the context of \cite{MNPS91} as it is solvable.  However, in our context,
  execution does more. To be specific, execution proceeds as follows: the system 
 requests the user to select a particular number for $x$. 
After the number -- 
 say, $5$ -- is selected, the system returns $y = 125$.
 
     As seen from the example above, universally quantified goals in intuitionistic logic
can be used to model 
interactive  tasks.

In this paper we present the syntax and semantics of this  language. 
The remainder of this paper is structured as follows. We describe  $fohh$
  logic  in
the next section.  Section 3 describes the new semantics.
Section 4 concludes the paper. 

\section{First-Order Hereditary Harrop Formulas}\label{sec:logic1}

The extended language is a version of Horn clauses
 with some extensions. It is described
by $G$- and $D$-formulas given by the syntax rules below:
\begin{exmple}
\>$G ::=$ \>  $A \sep   G \land  G \sep    \some x\ G \sep  \all x\ G \sep  D \supset G $ \\   \\
\>$D ::=$ \>  $A  \sep G \supset A\ \sep \all x\ D $\\
\end{exmple}
\noindent
In the rules above, $A$  represents an atomic formula.
A $D$-formula  is called a $fohh$.

In the transition system to be considered, $G$-formulas will function as 
queries and a set of $D$-formulas will constitute  a program. 
 We will  present the standard operational 
semantics for this language  as inference rules \cite{Khan87}. 
The rules for executing queries in our language are based on
uniform provability \cite{MNPS91}. Below the notation $D;\Pscr$ denotes
$\{ D \} \cup \Pscr$ but with the $D$ formula being distinguished
(marked for backchaining). Note that execution  alternates between 
two phases: the goal reduction phase (one  without a distinguished clause)
and the backchaining phase (one with a distinguished clause).

\begin{defn}\label{def:semantics}
Let $G$ be a goal and let $\Pscr$ be a program.
Then the task of proving $G$ from $\Pscr$ -- $\pr(\Pscr,G)$ -- is defined as follows:

\begin{numberedlist}

\item  $\pr(A;\Pscr,A)$. \% This is a success.

\item    $\pr((G_0\supset A);\Pscr,A)$ if 
 $\pr(\Pscr, G_0)$. \% backchaining

\item    $\pr(\all x D;\Pscr,A)$ if   $\pr([t/x]D;
\Pscr, A)$. 

\item    $\pr(\Pscr,A)$ if   $D \in \Pscr$ and $\pr(D;\Pscr, A)$. \% solving an atomic goal

\item $\pr(\Pscr, G_0 \land G_1)$  if $\pr(\Pscr, G_0)$ and 
$\pr(\Pscr, G_1)$.

 \item $\pr(\Pscr, D \supset G)$  if $\pr(\{ D \} \cup \Pscr, G)$.

 \item $\pr(\Pscr, \all x G)$  if $\pr(\Pscr, [y/x]G_1)$ where $y$ is a new free variable.

 \item $\pr(\Pscr, \some x G)$  if $\pr(\Pscr, [t/x]G)$ where $t$ is a constant or a variable.

\end{numberedlist}
\end{defn}
\noindent

\section{An Alternative Operational Semantics}\label{sec:0627}

Adding game semantics  requires fundamental changes to the execution model.

To be precise, our new execution model -- adapted from \cite{Jap03} -- now requires two phases:

\begin{numberedlist}

\item the proof phase: This phase builds a $proof\ tree$. This proof tree 
 encodes all the possible execution sequences.

\item the execution phase: This phase actually solves the goal relative to the program using the proof tree.

\end{numberedlist}

\noindent 

Note that a proof tree can be represented 
as a list and this idea is used here.
Now, given a program $\Pscr$ and a goal $G$, a proof tree of $\Pscr \supset G$  is a list of tuples of
the form $\lb E,i \rb$ or $\lb E,(i,j) \rb$ where $E$ is a (proof) formula and $i,j$ are the distances to $F$'s chilren
in the proof tree. Below, $a_1::\ldots::a_n::nil$ represents a list of $n$ elements.

\begin{defn}\label{def:semantics}
Let $G$ be a goal and let $\Pscr$ be a program.
Then the task of proving $\Pscr\supset G$ and returning its proof tree $L$ -- 
written as $\pr(\Pscr\supset G,L)$ -- is defined as follows:

\begin{numberedlist}

\item  $\pr(E,\lb E, - \rb::nil)$
       if $E = A;\Pscr\supset A$. \% This is a leaf node.

\item    $\pr(E, \lb E,1\rb::L)$ if 
 $E = (G_0\supset A);\Pscr\supset A$     and                  $\pr(\Pscr\supset G_0,L)$.

\item    $\pr(E, \lb E,1\rb::L) $ if  
       $E =  \all x D;\Pscr\supset A$ and                 $\pr([t/x]D;\Pscr\supset A,L) $.

\item    $\pr(E,\lb E,1\rb::L)$ if  
 $E = \Pscr\supset A$ and   $D \in \Pscr$ and $\pr(D;\Pscr\supset A,L)$.

\item $\pr(E,\lb E,(m+1,1) \rb::L_2 )$ 
 if  $E = \Pscr\supset (G_0 \land G_1)$ and $\pr(\Pscr\supset G_0,L_0)$ and 
$\pr(\Pscr\supset G_1,L_1)$ and $append(L_0,L_1,L_2)$ and  $length(L_1,m)$. 

\item $\pr(E,\lb E,1\rb::L)$  if
            $E = \Pscr\supset (D\supset  G)$ and   $\pr((\{ D \}\cup\Pscr)\supset G ,L)$
  
 \item $\pr(E,\lb E,1\rb::L)$  if
 $E =  \Pscr\supset \some x G$ and                $\pr(\Pscr\supset [t/x]G ,L)$
where $t$ is a constant or a variable.

 \item $\pr(E,\lb E,1\rb::L)$  if 
 $E =  \Pscr\supset \all x G$ and    $\pr(\Pscr\supset [y/x]G ,L)$
where $y$ is a new free variable.

\end{numberedlist}
\end{defn}
\noindent

Once a proof tree is built, the execution phase actually solves the goal relative to the program 
using the proof tree. In addition, to deal with the universally quantified goals properly,
the execution needs to maintain an $input$ $substitution$ of the form
$\{ y_0/c_0,\ldots,y_n/c_n \}$ where each $y_i$ is a variable introduced by a universally quantified goal
in the proof phase and each $c_i$ is a constant typed by
the user during the execution phase.

\begin{defn}\label{def:exec}
Let $i$ be an index and let $L$ be a proof tree and $F$ is an input substitution.  
Then   executing $L_i$ (the $i$ element in $L$) with $F$ -- written as $\prove(i,L,F)$ --
 is defined as follows: 

\begin{numberedlist}

\item  $\prove(i,L,F)$ if $L_i = (E,-)$. \% no child

\item  $\prove(i,L,F)$ if $L_i = (\Pscr\supset  G_0 \land G_1, (n,m))$ and
 $\prove(i-n,L,F)$ and $\prove(i-m,L,F)$. \% two children

\item  $\prove(i,L,F)$ if $L_i = (\Pscr\supset \all x G, 1)$ and $L_{i-1} = (\Pscr\supset [y/x] G, n)$
  and $read(k)$ and $\prove(i-1,L, F \cup \{ y/c \})$ where $c$ is the user input (the value 
stored in $k$). 

\item  $\prove(i,L,F)$ if $L_i = (\Pscr\supset \some x G, 1)$ and $L_{i-1} = (\Pscr\supset [t/x] G, n)$
  and $\prove(i-1,L',F)$ where   $c = F(t)$ and $L'$ is identical to $L$ except that 
$L'_{i-1} = (\Pscr\supset [c/x] G, n)$.  Hence the term $t$ must be replaced by $c$ to ensure correct
operation.

\item  $\prove(i,L,F)$ if $L_i = (\all x D;\Pscr\supset A, 1)$  and $L_{i-1} = ([t/x]D;
\Pscr\supset A, n)$
  and $\prove(i-1,L',F)$ where   $c = F(t)$ and $L'$ is identical to $L$ except that 
$L'_{i-1} = ([c/x]D;\Pscr\supset A, n)$. Hence the term $t$ must be replaced by $c$ to ensure correct
operation.

\item  $\prove(i,L,F)$ if $L_i = (E,1)$  and $\prove(i-1,L)$. \% otherwise

\end{numberedlist}
\end{defn}

\noindent  
Now given a program $\Pscr$ and a goal $G$, $L$ is initialized to the proof tree of  $\Pscr \supset G$,
and $F$ is initialized to an empty substitution and $n$ is initialized to the length of $L$.

\section{Conclusion}\label{sec:conc}

In this paper, we have considered a new execution model for $fohh$.
 This new model is interesting in that it gives a logical status to the $read$ predicate in Prolog.
We plan to connect our execution model to Japaridze's Computability Logic \cite{Jap03,Jap08}
 in the near future.



\bibliographystyle{plain}



\end{document}